\documentclass[showpacs,epsf,twocolumn,superscriptaddress]{revtex4-1}
\usepackage{float}
\usepackage{amsmath}
\usepackage{color,graphicx}
\usepackage{hyperref}

\begin{document}

\title{Complex exchange mechanism driven ferromagnetism in half-metallic Heusler Co$_{2}$TiGe: Evidence from critical behavior}

\author{Shubhankar Roy}
\affiliation{Saha Institute of Nuclear Physics, HBNI, 1/AF Bidhannagar, Kolkata 700 064, India}
\affiliation{Vidyasagar Evening College, 39, Sankar Ghosh Lane, Kolkata 700 006, India}
\author{Nazir Khan}
\affiliation{Saha Institute of Nuclear Physics, HBNI, 1/AF Bidhannagar, Kolkata 700 064, India}
\author{Ratnadwip Singha}
\affiliation{Saha Institute of Nuclear Physics, HBNI, 1/AF Bidhannagar, Kolkata 700 064, India}
\author{Arnab Pariari}
\affiliation{Saha Institute of Nuclear Physics, HBNI, 1/AF Bidhannagar, Kolkata 700 064, India}
\author{Prabhat Mandal}
\affiliation{Saha Institute of Nuclear Physics, HBNI, 1/AF Bidhannagar, Kolkata 700 064, India}
\date{\today}

\begin{abstract}
We have investigated the critical phenomenon associated with the magnetic phase transition in the half-metallic full-Heusler Co$_2$TiGe. The compound undergoes a continuous ferromagnetic to paramagnetic phase transition at the Curie temperature $T_{C}$=371.5 K. The analysis of magnetization isotherms in the vicinity of $T_{c}$, following modified Arrott plot method, Kouvel-Fisher technique, and critical isotherm plot, yields the asymptotic critical exponents $\beta$=0.495, $\gamma$=1.324, and $\delta$=3.67. The self-consistency and reliability of the obtained exponents are further verified by the Widom scaling relation and scaling equation of states. The mean-field-like value of the critical exponent $\beta$ suggests long-range nature of the exchange interactions, whereas the values of the critical exponents $\gamma$ and $\delta$, imply sizeable critical spin fluctuations. The half-metallic itinerant character of Co$_{2}$TiGe in the presence of magnetic inhomogeneity may result in such a strong deviation from the three-dimensional Heisenberg values ($\beta$=0.369, $\gamma$=1.38 and $\delta$=4.8) of the critical exponents towards the mean field values ($\beta$=0.5, $\gamma$=1 and $\delta$=3). The results suggest complex nature of exchange couplings that stabilize the long-range ferromagnetic ordering in the system and are consistent with the earlier theoretical studies on the exchange mechanism in Co$_2$TiGe.
\end{abstract}
\maketitle

\section{Introduction}

The discovery of ferromagnetism by Friedrich Heusler in Cu$_2$MnAl, which does not contain any of the pure ferromagnetic elements, initiated the evergrowing research interests in the inter-metallic compounds \cite{Heusler}. Subsequently, numerous experimental works \cite{Webster1,Webster2,Hamzic} have been carried out and several theoretical models \cite{Price,Kubler,JLMor,Kurtulus,Sasioglu,Thoene,Trudel} have been proposed to understand the microscopic interactions that lead to the long-range magnetic ordering in this class of materials. The fascinating half-metallic character \cite{Groot,Galanakis} makes them very promising candidates for the spintronics applications at room temperature \cite{Chappert}. Moreover, recently, novel topological semi-metal states \cite{Chang} have been discovered in these systems. The Co$_2$-based ternary intermetallic Heusler compounds viz., Co$_2$$M'$$Z$, where $M'$ is a transition metal and $Z$ is a main group element, received special interest and became the most widely studied systems because of their very high Curie temperatures \cite{Trudel}.

Co$_2$TiGe crystallizes in cubic $L$2$_1$ structure (space group: $Fm$-$3m$), which consists of four inter-penetrating face-centered-cubic (fcc) lattices. The crystallographic positions of Co atoms are (0,0,0) and (1/2, 1/2, 1/2) and that of the Ti and Ge atoms are (1/4, 1/4, 1/4) and (3/4, 3/4, 3/4), respectively. There are several theoretical and experimental reports exploring the structural, electronic, and magnetic properties of Co$_2$TiGe \cite{Barth}. The spin-resolved band structure calculations show that the majority spin-band has a metallic character, whereas the minority spin-band exhibits semiconducting behavior with a band gap of about 0.5 eV around the Fermi level \cite{Chang,Barth}. Further, the inclusion of the spin-orbit coupling (SOC), albeit very weak, in the band structure calculation shows novel topological Weyl semimetal state \cite{Chang}. The temperature dependence of the resistivity measurement in Co$_2$TiGe shows typical metallic behavior down to 2 K \cite{Barth}. The origin of ferromagnetism in half-metallic Co$_2$-based Heusler alloys is a rather complicated issue and remains one of the most interesting problems in modern magnetism. They exhibit Slater-Pauling-type behavior for the magnetization, where the saturation magnetization scales with the total number of valance electrons in the unit cell, leading to a characteristic integer magnetic moments. Magnetic Heusler alloys are traditionally believed to be ideal local-moment systems \cite{Telling} and their exchange couplings can be described by a Heisenberg Hamiltonian. However, several theoretical reports claim the presence of complex exchange interactions of the localized magnetic moments in the Co$_2$-based Heusler compounds \cite{Kubler,Kurtulus,Sasioglu,Thoene,Trudel}. Both the short-range direct exchange coupling between the nearest neighbor spins and the coupling between further neighbor spins mediated by the Ruderman-Kittel-Kasuya-Yosida (RKKY) type long-range interaction, are believed to be responsible for the long-range magnetic ordering \cite{Trudel}. It is concluded that inter-sublattice and intra-sublattice exchange interactions of different strength stabilize robust ferromagnetic ordering with remarkably high Curie temperatures in the Co$_2$-based Heusler compounds. However, a recent theoretical study shows that in the Co$_2$Ti$Z$ system, the magnetic moment of the Co atoms and that of the Ti atoms are not completely localized, which is further supported by the disorder local-moment calculation \cite{Barth}. Thus, the presence of localized magnetic moment is not prerequisite for the occurrence of half-metallic ferromagnetism. Furthermore, significant orbital magnetic moments in different Co$_2$$M'$$Z$ compounds as probed by x-ray magnetic circular dichroism studies \cite{Telling,Klaer,Miyamoto}, indicate that the SOC may result in magneto-crystalline anisotropy. So, anisotropic exchange interaction may also be involved in the magnetic ordering phenomenon in these systems. The values of the critical exponents associated with a continuous magnetic phase transition, reflect the nature of the exchange mechanism, spin symmetry or the magnetic anisotropy, and the effective dimensionality of spin-spin interactions in the system \cite{Fisher}. Therefore, by experimentally investigating the critical phenomenon associated with the ferromagnetic to paramagnetic phase (FM-to-PM) transition in Co$_2$TiGe, we elucidate the nature of exchange mechanism in a Co$_2$-based Heusler alloy.

\section{Experimental details}

Co$_{2}$TiGe compound was prepared by arc melting of stoichiometric amounts of its constituents in a highly purified argon atmosphere. The obtained ingots were then sealed in an evacuated quartz tube and annealed for about 3 weeks. Phase purity and the structural analysis of the sample were done using the powder x-ray diffraction (XRD) technique with Cu-K$_{\alpha}$ radiation in a Rigaku x-ray diffractometer (TTRAX III). The Rietveld profile refinement of the XRD pattern (Fig. 1) shows that the compound is single phase in nature with space group symmetry $L$2$_1$. The refined lattice parameters, $a$=$b$=$c$=5.818(2) {\AA}, are in good agreement with that reported previously \cite{Barth,Barth2}. The magnetization measurements were done using a Superconducting Quantum Interference Device-Vibrating Sample Magnetometer (SQUID-VSM) (MPMS 3, Quantum Design) in fields up to 5 T. The sample used for the magnetic measurements is of approximate dimensions $0.5\times0.5\times4$ mm$^{3}$. To minimize the demagnetization effect the external magnetic field was applied along the longest sample direction. The magnetization data were taken over the temperature range from 366 to 377 K at 1.0 K interval. To achieve good thermal equilibrium, we have stabilized each temperature for 45 minutes. For each M(H) isotherm, the magnetic field was increased from 0 to 5T and then  reduced to 0. We didn't find any difference in $M(H)$ between the increasing and decreasing field.

\begin{figure}
\includegraphics[width=0.5\textwidth]{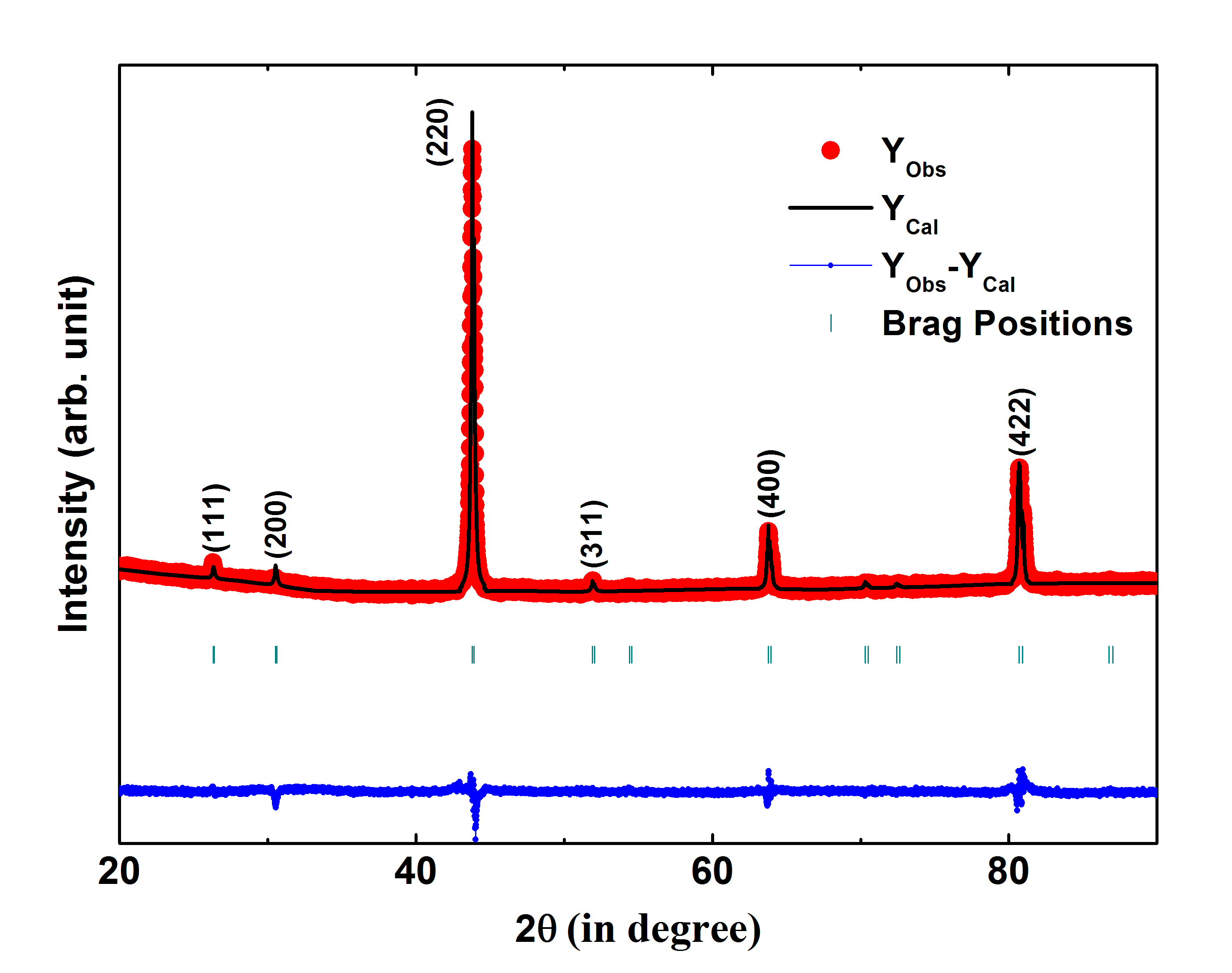}
\caption{(Color online) X-ray diffraction pattern of powdered samples of Co$_{2}$TiGe. Red circles are experimental data (Y$_{obs}$), black line is the calculated pattern (Y$_{cal}$), blue line is the difference between experimental and calculated intensities (Y$_{obs}$-Y$_{cal}$), and green vertical lines show the Bragg positions.}\label{rh}
\end{figure}

\section{Results and Discussions}

Figure 2 shows the magnetic field dependence of dc-magnetization ($M$) at 2 K. In the inset, we have plotted $M$ as a function of temperature, measured at 500 Oe. The observed behavior is quite similar to the earlier reports \cite{Barth,Barth2}. As mentioned before, the Co-based Heusler alloys, which are half-metallic ferromagnets (FMs), exhibit the Slater-Pauling-type behavior of the magnetization as given by $M_{p}=(Z_{p}-24)$ $\mu_{B}$/f.u. Here, $M_{p}$ is the total magnetic moment and $Z_{p}$ is the total number of valence electrons in the unit cell of the compound. For Co$_{2}$TiGe, the value of $Z_{p}$ is 26. Therefore, according to the above relation, the total magnetic moment should be 2$\mu_{B}$/f.u. From  Fig. 2, the saturation magnetization is estimated to be $\sim1.99$ $\mu_{B}$/f.u., which implies that the compound obeys the Slater-Pauling rule. As shown in the inset of Fig. 2, the magnetization curve mimics a continuous or second order FM-to-PM phase transition. From the temperature derivative of this magnetization curve, the Curie temperature ($T_{C}$) is estimated to be 371 K.

\begin{figure}
\includegraphics[width=0.5\textwidth]{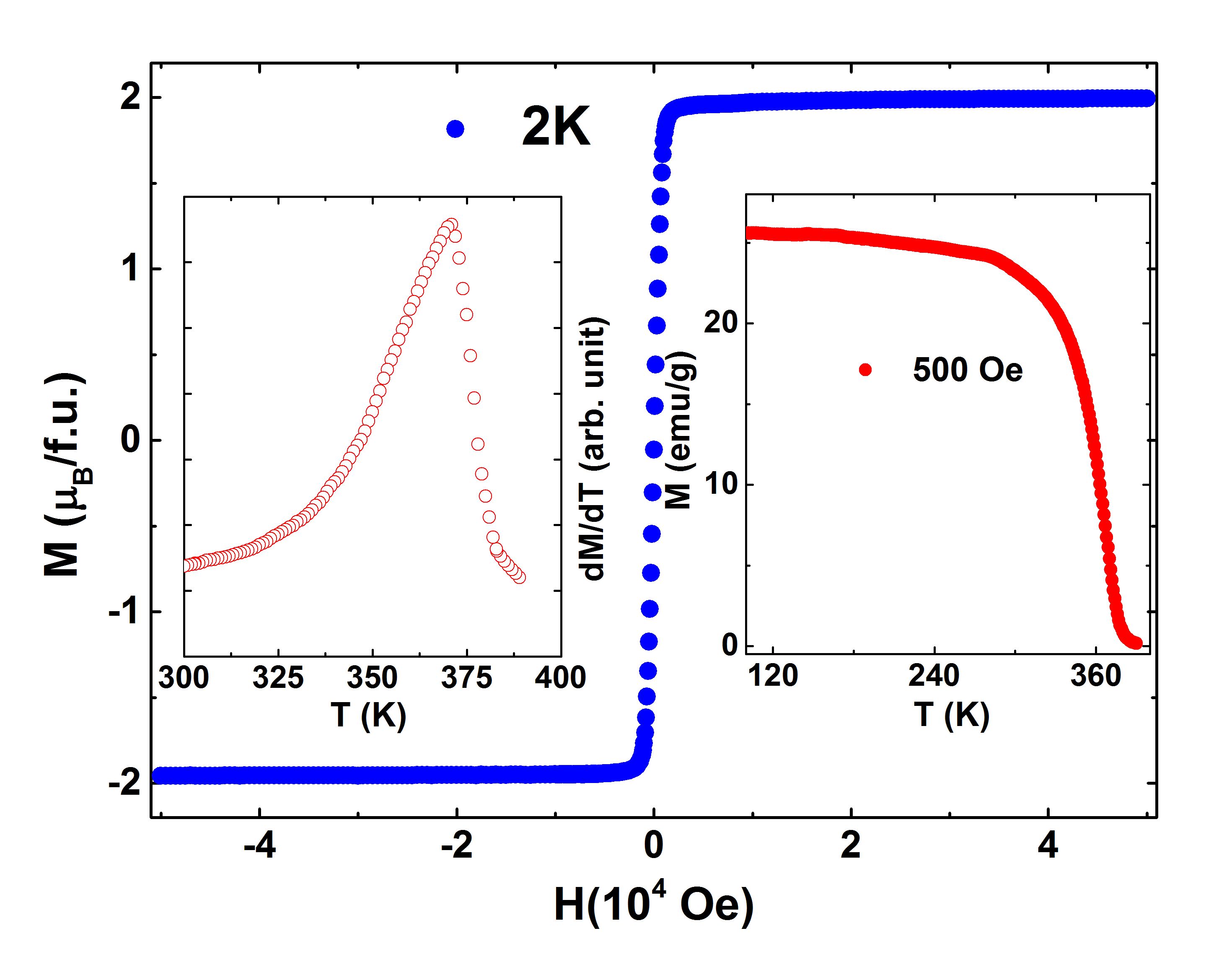}
\caption{(Color online) Field dependence of the magnetization ($M$ vs. $H$) at 2 K. Left inset: dM/dT versus temperature. Right inset: temperature dependence of the magnetization ($M$ vs. $T$) measured at 500 Oe. }\label{rh}
\end{figure}

A continuous magnetic phase transition exhibits critical behavior of different thermodynamic variables, governed by the critical fluctuation. The critical behavior is characterized by a set of static critical exponents. The values of these exponents depend only on the symmetry of the order parameter and the lattice dimensionality of the system, resulting in different universality classes of the continuous phase transition in uniform FMs.  In the vicinity of $T_{C}$, the spin-spin correlation length ($\xi$) diverges as, $\xi=\xi_0|(T-T_{C})/T_{C}|^{-\nu}$, which leads to universal scaling laws for the spontaneous magnetization, $M_{S}$(0, $T$), and inverse initial susceptibility, ${\chi_0}^{-1}$(0, $T$). $M_{S}(0, T)$ below $T_{C}$, ${\chi_0}^{-1}(0, T)$ above $T_{C}$, and the magnetization isotherm $M(H, T_C)$ at $T_{C}$ are characterized by a set of static critical exponents $\beta$, $\gamma$, and $\delta$, respectively. These exponents are defined as follows \cite{Eugene}:

\begin{equation}
M_S(0, T) =\left[M_S(0)\right](-\varepsilon)^{\beta}, \varepsilon< 0,
\end{equation}

\begin{equation}
{\chi_0}^{-1}(0, T) = \left[\frac{H_0}{M_S(0)}\right]{(\varepsilon)}^{\gamma}, \varepsilon> 0,
\end{equation}

\begin{equation}
M(H, T_C) = D(H)^{1/\delta}, \varepsilon = 0,
\end{equation}

where $\varepsilon=\frac{T-T_{C}}{T_{C}}$ is the reduced temperature and $M_S(0)$, $\frac{H_0}{M_S(0)}$, and $D$ are the critical amplitudes. The scaling theory predicts that in the close vicinity of the phase transition, the magnetic equation of state for the system can be expressed as
\begin{equation}\label{scaling}
M(H,\varepsilon) = |\varepsilon|^{\beta}
f_{\pm}\left[\frac{H}{|\varepsilon|^{(\gamma + \beta)}}\right],
\end{equation}
where $f_{+}$ and $f_{-}$ are regular functions for temperature above and below $T_{C}$, respectively. In terms of the renormalized magnetization, $m$$\equiv$$\mid\varepsilon\mid$$^{-\beta}$$M(H, \varepsilon)$, and the renormalized field, $h$$\equiv$$\mid$$\varepsilon$$\mid$$^{-(\gamma + \beta)}$$H$, the above scaling equation can be rewritten as
\begin{equation}
m =f_{\pm}\left(h\right).
\end{equation}
The above equation implies that for the right choice of $\beta$, $\gamma$, and $\varepsilon$, the $m$ vs. $h$ isotherms will fall onto two separate branches of the scaling function: $f_{+}$ for isotherms above $T_{C}$ and $f_{-}$ for isotherms below $T_{C}$. This is also an important criterion to check whether the set of the critical exponents are the same below and above $T_{C}$.

To characterize the FM-to-PM phase transition in the present system, a series of magnetization isotherms, as shown in Fig. 3(a), has been measured in the vicinity of the approximate Curie temperature, which was determined from the inset of Fig. 2. To test whether the framework of the Landau mean-field theory of magnetic phase transition holds for Co$_2$TiGe, we have constructed the conventional Arrott plot \cite{Arrott} of $M^{2}$ vs. $H/M$ isotherms as shown in Fig. 3(b). For the mean-field values of the critical exponents ($\beta=0.5$ and $\gamma=1$), Arrott plot should generate a set of parallel straight lines and the one that passes through the origin of the plot, corresponds to the isotherm at the exact critical temperature. However, the Arrott plot for Co$_2$TiGe does not constitute a set of parallel straight lines and shows significant non-linearity with downward curvature even at high magnetic fields. Therefore, the Landau mean-field theory fails to describe the phase transition in Co$_2$TiGe and suggests the presence of significant critical fluctuations. According to the Banerjee's criterion \cite{Banerjee}, the positive values of the slope of the curves in the Arrott plot, indicate second order nature of the phase transition. In Fig. 3(b), all the $M^{2}$ vs. $H/M$ isotherms show concave downward curvatures, which yield positive values of the slopes and, thereby, confirm second-order FM-to-PM phase transition in Co$_2$TiGe.

\begin{figure}
\includegraphics[width=0.5\textwidth]{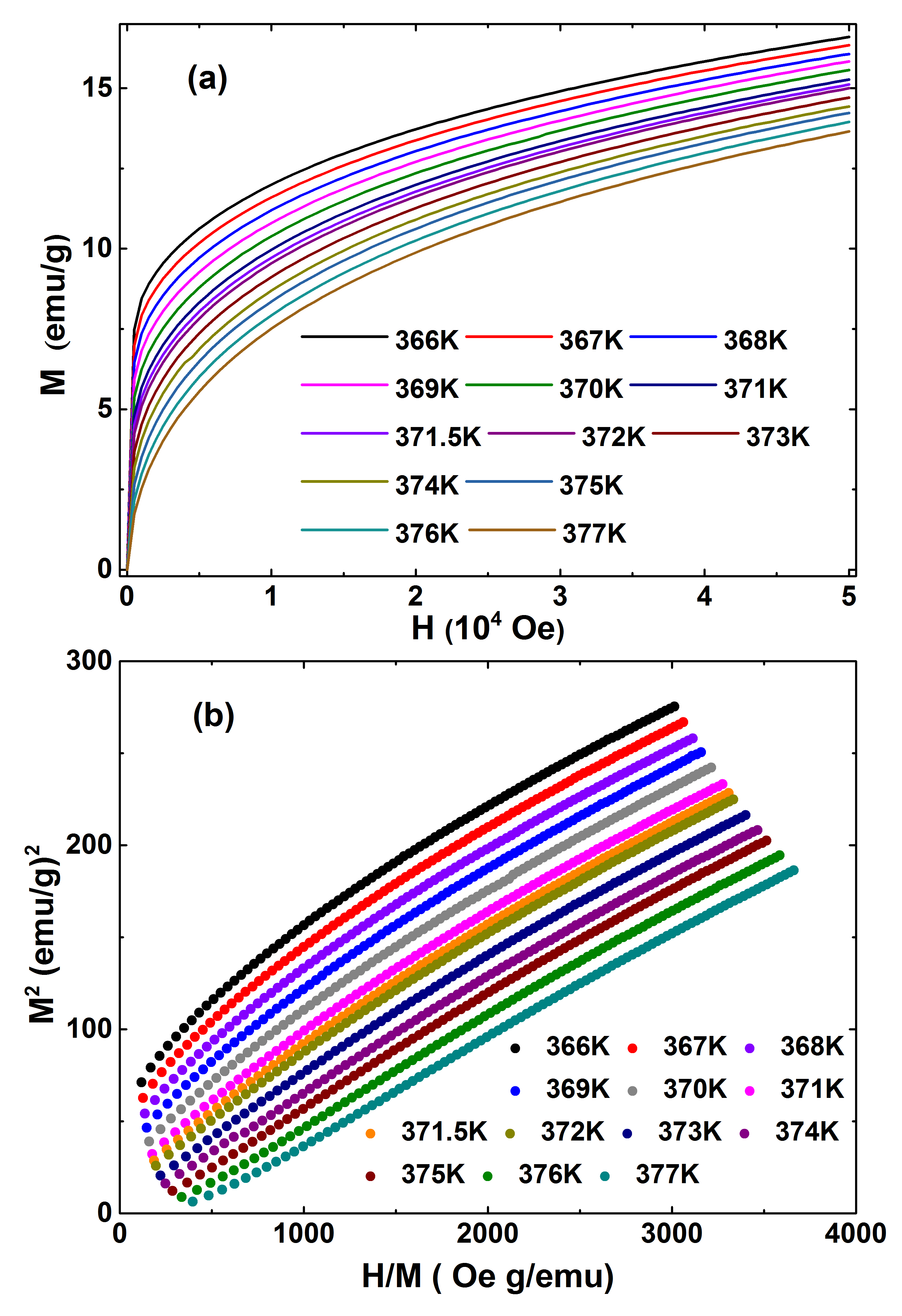}
\caption{(Color online) (a) Isothermal magnetisation ($M$ vs. $H$) as a function of magnetic field measured around $T_{C}$. (b) Arrott plot of isotherms measured in the vicinity of  $T_{C}$.}\label{rh}
\end{figure}

To determine the true critical exponents, we have reanalyzed the magnetization isotherms using the modified Arrott plot \cite{Arrott2} method, which is based on the following Arrott-Noaks equation of state:
\begin{equation}
(H/M)^{1/\gamma}=a\left(\frac{T-T_{C}}{T_{C}}\right)+bM^{1/\beta}.
\end{equation}

\begin{figure*}
\includegraphics[width=1.0\textwidth]{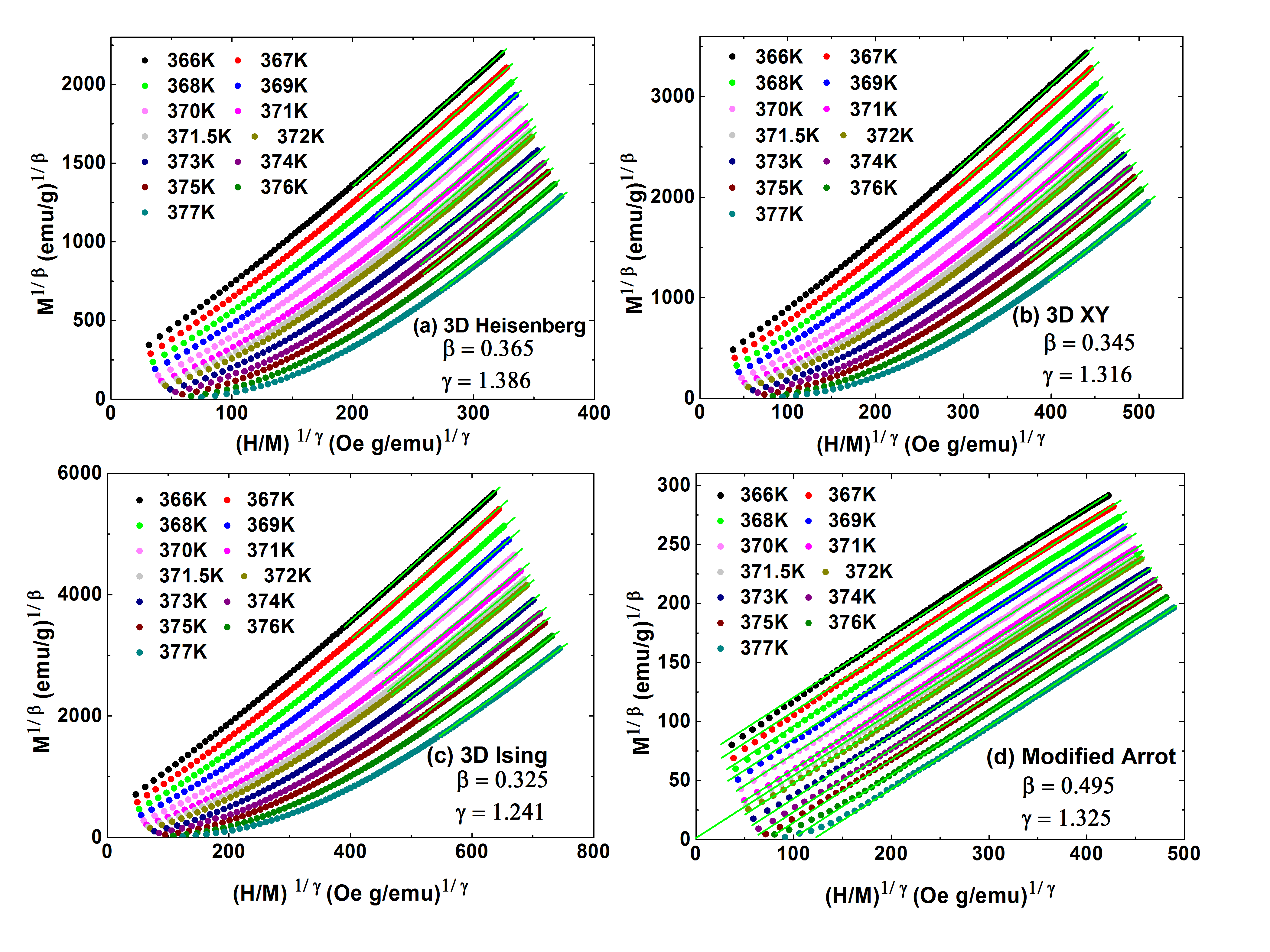}
\caption{(Color online) Modified Arrott plot of isotherms for 366 K$\leq$ $T$ $\leq$377 K, with the parameters of (a) 3D Heisenberg model, (b) 3D XY model, and (c) 3D Ising model. (d) Modified Arrott plot of isotherms with $\beta=0.495$ and $\gamma=1.325$. Solid lines are the linear fits to the isotherms in the high-field regions.}\label{rh}
\end{figure*}

\begin{figure}
\includegraphics[width=0.5\textwidth]{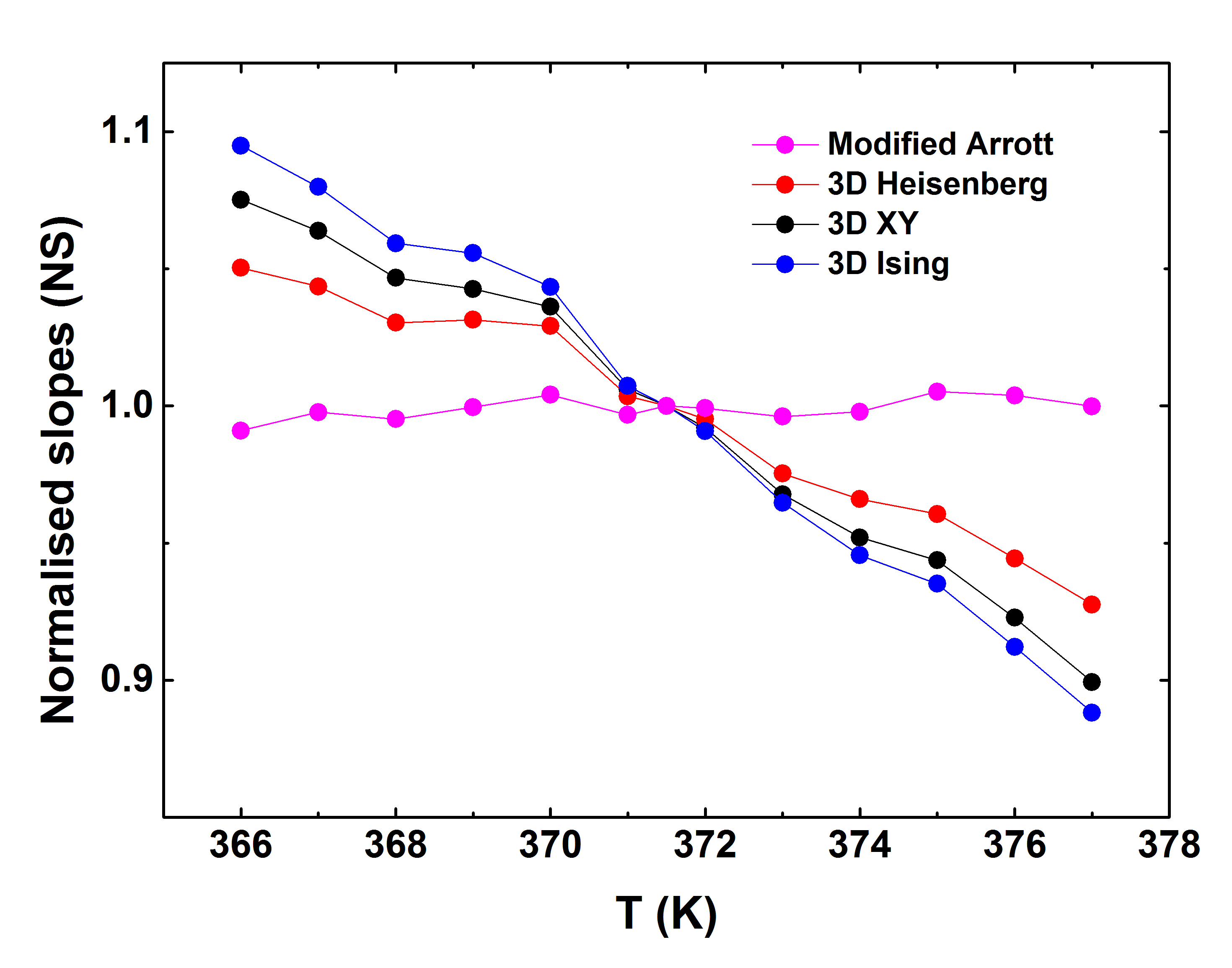}
\caption{(Color online) Temperature dependence of the normalized slopes, $NS=S(T)/S(T_C$), obtained from Figs. 4(a)-(d).}\label{rh}
\end{figure}

Here, $a$ and $b$ are constants. The critical exponents predicted for three-dimensional (3D) system by theoretical models such as 3D Heisenberg ($\beta= 0.365$, $\gamma= 1.386$), 3D XY ($\beta= 0.345$, $\gamma= 1.316$), and 3D Ising model ($\beta= 0.325$, $\gamma= 1.24$) have been used to construct modified Arrott plots as shown in Figs. 4(a), 4(b), and 4(c), respectively. We can see from Figs. 4(a)-4(c) that at the high-field region the constructions show quasi-straight lines, which appear to be parallel to each other. To determine which one among these three models generates the best modified Arrott plot with parallel isotherms, we have compared the slope of these straight lines (obtained from the linear fits to the isotherms at high-field region) to that obtained for the isotherm at $T \simeq T_C$, for each modified Arrott plot. Here, the slope is defined as $S(T)=dM^{1/\beta}/d(H/M)^{1/\gamma}$. The obtained normalized slopes, $NS = S(T )/S(T_{C})$, for each model have been plotted as a function of temperature in Fig. 5. It is evident that the values of the $NS$ deviate progressively from the ideal value of 1 both below and above $T_C$. However, among these three models, the deviation is minimum for the 3D Heisenberg model. Therefore, the modified Arrott plot corresponding to 3D Heisenberg model [Fig. 4(a)] has been used for further refinement of the exponents $\beta$ and $\gamma$ using a rigorous iterative method as described below.

In Fig. 4(a), the intercepts of the linear fits (to the isotherms in high-field region) with the axes $M^{{1/\beta}}$ and ($H/M$)$^{1/\gamma}$, give reliable values of $M_S(0,T)$ and ${\chi_0}^{-1}(0,T)$, respectively. The power law fits to the data following Eqs. (1) and (2), yield new values of the exponents $\beta$ and $\gamma$, which are used to construct a new modified Arrott plot. During the fitting, $T_{C}$ was varied to get the best fit results. From this plot, new $M^{{1/\beta}}$ and ($H/M$)$^{1/\gamma}$ are obtained by linear fits to the isotherms at high-field region. So we get a new set of $\beta$ and $\gamma$. This procedure is repeated to achieve self-consistency, i.e., when the new values of the critical exponents become almost equal to the values before that iteration.

The final modified Arrott plot, constructed using $\beta$=0.495 and $\gamma$=1.325, is shown in Fig. 4(d). The normalized slopes of the linear fits to the isotherms of this modified Arrott plot, are also shown in Fig. 5. Here, one can see that both above and below $T_{C}$, $NS$ is close to 1, implying that all the isotherms form parallel straight lines with the isotherm at $T$=371.5 K, which almost passes through the origin of the plot. Thus, the more accurate value of $T_{C}$ is estimated to be 371.5 K. In the low-field region, the small curvatures in the isotherms of Fig. 4(d) is due to the reorientation of the domains at low magnetic fields. The temperature dependence of $M_S(0,T)$ and ${\chi_0}^{-1}(0,T)$, which are obtained from Fig. 4(d), are shown in Fig. 6(a). The power law fit following Eq. (1) to the $M_S(0,T)$ data yields $\beta$=0.494(3) and $T_C$=371.5(1) and fitting of Eq. (2) to the ${\chi_0}^{-1}(0,T)$ data gives $\gamma$=1.33(3) and $T_C$=371.5(1). Theses values of the exponents and $T_{C}$, are almost same to that obtained from the modified Arrott plot [Fig. 4(d)]. Therefore, the estimated values of $\beta$, $\gamma$, and $T_C$, are very reliable and intrinsic.

\begin{figure}
\includegraphics[width=0.5\textwidth]{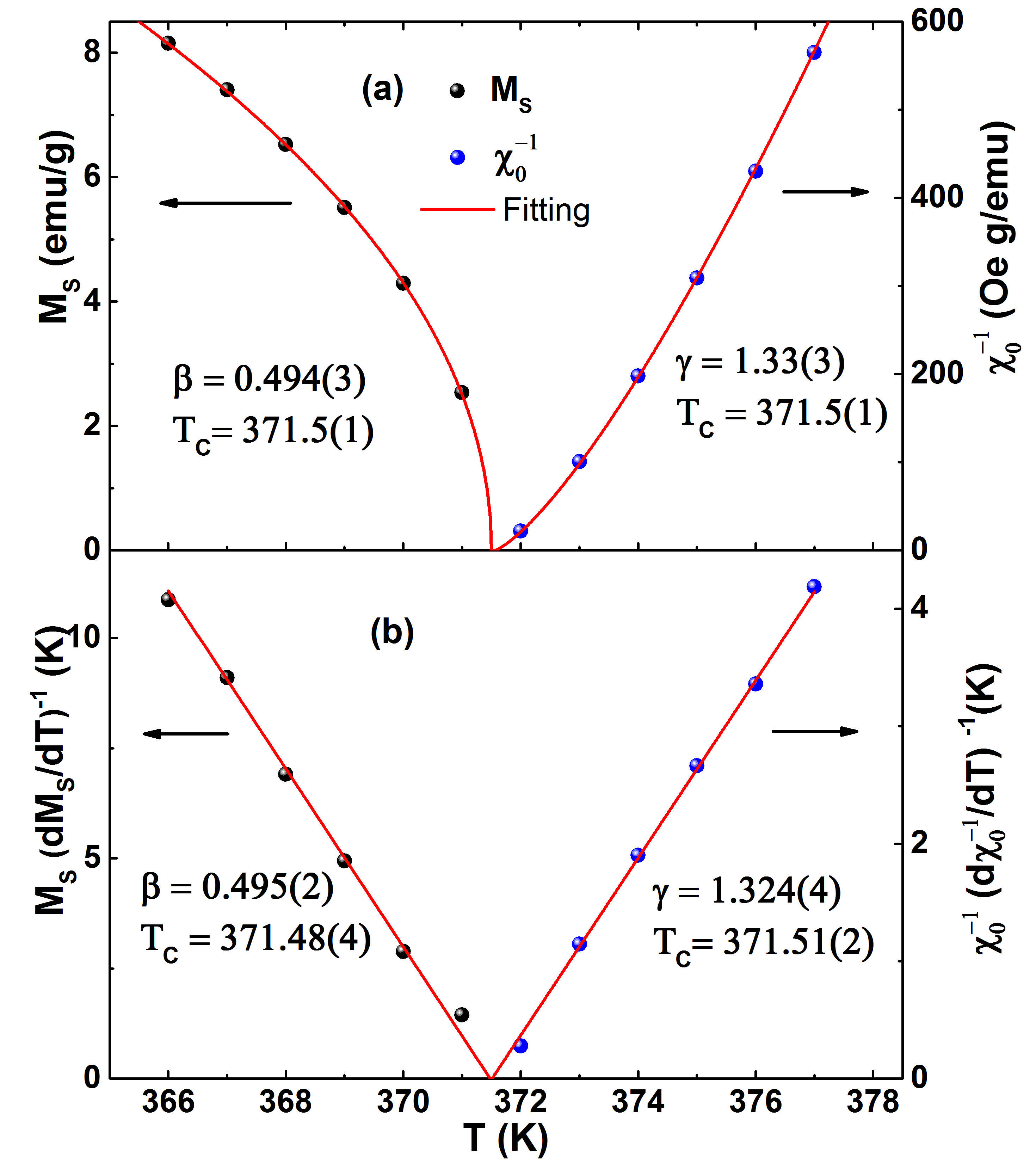}
\caption{(Color online) (a) Temperature variations of $M_{S}$ and ${\chi_0}^{-1}$ along with the fits (solid lines) following Eqs. (1) and (2), which give the values of the exponents and $T_{C}$ as mentioned in the plot. (b) Kouvel-Fisher plot of M$_{S}$ and ${\chi_0}^{-1}$. The exponents and $T_{C}$ are obtained from the linear fits (solid lines) of the data.}\label{rh}
\end{figure}

The accurate values of the exponents as well as $T_{C}$ can also be obtained using the Kouvel-Fisher technique \cite{Kouvel}. According to this method, $M_{S}(\frac{dM_{S}}{dT})^{-1}$ vs. $T$ and $[{\chi_0}\frac{d{\chi_0}^{-1}}{dT}]^{-1}$ vs. $T$ should be straight lines with slope $\frac{1}{\beta}$ and $\frac{1}{\gamma}$, respectively, and their intercepts on the $T$-axis give the value of $T_{C}$. The linear fits to these plots [shown in Fig. 6(b)] yield $\beta$=0.495(2), $T_{C}$=371.48(4) and $\gamma$=1.324(4), $T_{C}$=371.51(2).

The critical exponent $\delta$ can be obtained from the critical isotherms i.e., the $M(H)$ isotherm at $T=T_C$ as well as from the Widom scaling relation \cite{widom}:
\begin{equation}
\delta=1+\frac{\gamma}{\beta}.
\end{equation}

\begin{figure}
\includegraphics[width=0.5\textwidth]{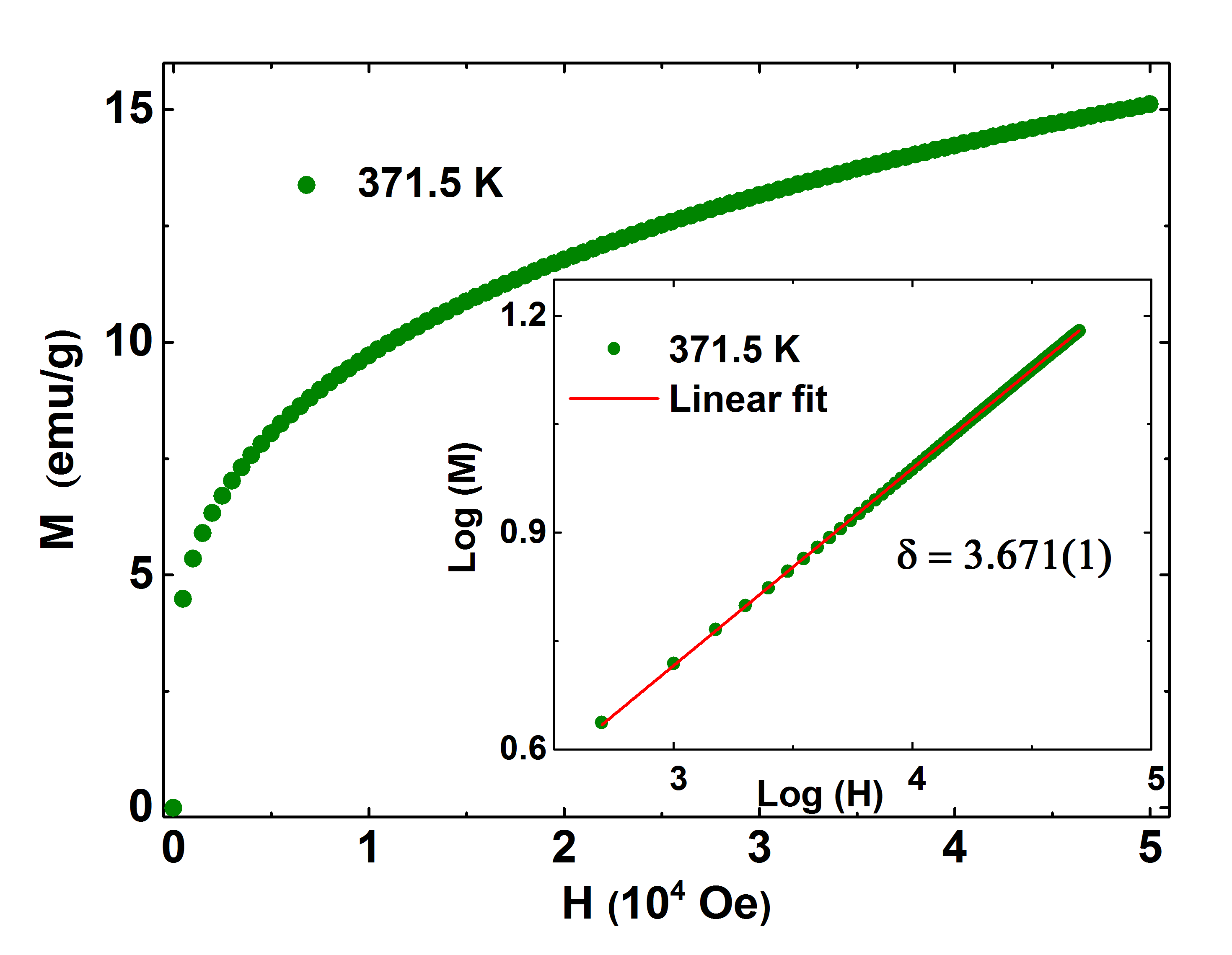}
\caption{(Color online) Field dependence of the magnetization isotherm at $T_C$=371.5 K for Co$_2$TiGe. The inset shows the same plot in log-log scale, where the solid line is the linear fit following Eq. (3) that gives the critical exponent $\delta$ as mentioned in the graph.}\label{rh}
\end{figure}

The $M(H)$ plot at $T$=371.5 K is shown in Fig. 7 and the inset shows the same plot on the log-log scale. Following Eq. (3), from the inverse of slope of the linear fit to the log($M$) vs. log($H$) curve, we obtain $\delta$=3.671(1). Using the values of $\beta$ and $\gamma$ obtained from Figs. 6(a) and 6(b), the Widom scaling relation, Eq. (7), gives $\delta=3.692$ and $\delta=3.675$, respectively. These values of $\delta$ are very close to that obtained from the critical isotherms (Fig. 7). Hence the obtained values of critical exponents and $T_{C}$ are self-consistent and obey the Widom scaling relation very well.

\begin{table*}
\caption{Comparison of critical exponents of Co$_{2}$TiGe with earlier reports on itinerant uniform FM Ni, itinerant amorphous FM Gd$_{80}$Au$_{20}$, and different theoretical models. Abbreviation: modified Arrott plot (MAP); Kouvel-Fisher (KF); critical isotherm (CI); renormalization group epsilon ($\epsilon^{'}$ = 2$\sigma$ - $d$) expansion(RG-$\epsilon^{'}$).}
\centering
\newcommand*{\TitleParbox}[1]{\parbox[c]{1.8cm}{\raggedright #1}}
\begin{tabular*}{1\textwidth}{@{\extracolsep{\fill}} c c c c c c c }
\hline \hline

Material/Model & Ref.&Technique &$T_{C}$ &$\beta$ &$\gamma$ &$\delta$\\[0.5pt]
\hline \hline
Co$_{2}$TiGe &This work &MAP &371.5  &0.495 &1.325  &3.677\\[6pt]
  &This work &KF&371.5(1)  &0.495(2) &1.324(4)  &3.675\\[6pt]
  &This work &CI    &      &      &      &3.671(1)\\[6pt]
Ni &\cite{kaul} &KF  &      & 0.391(10) & 1.314(16)  &4.39(2)\\[6pt]
Gd$_{80}$Au$_{20}$ &\cite{sjpoon} &KF&      &0.44(2)  &1.29(5)  &3.96(3)\\[6pt]
Mean Field &\cite{Eugene} &Theory  &      &0.5  &1  &3\\[6pt]
3D Heisenberg &\cite{Banerjee,Campostrini} &Theory  &      &0.365  &1.386  &4.8\\[6pt]
3D XY &\cite{Banerjee,Guillou} &Theory  &      &0.345  &1.316  &4.81\\[6pt]
3D Ising &\cite{Banerjee,Guillou} &Theory  &      &0.325  &1.24  &4.82\\[6pt]
LR exchange: $J(r)=1/r^{d+\sigma}$\\[0pt]
$d$ = 3, $n$ =3, $\sigma$= 1.88 &\cite{MEFisher} &RG-$\epsilon^{'}$  &      &0.393  &1.32  &4.36\\[6pt]

\hline \hline

\end{tabular*}

\end{table*}
\begin{figure}
\includegraphics[width=0.5\textwidth]{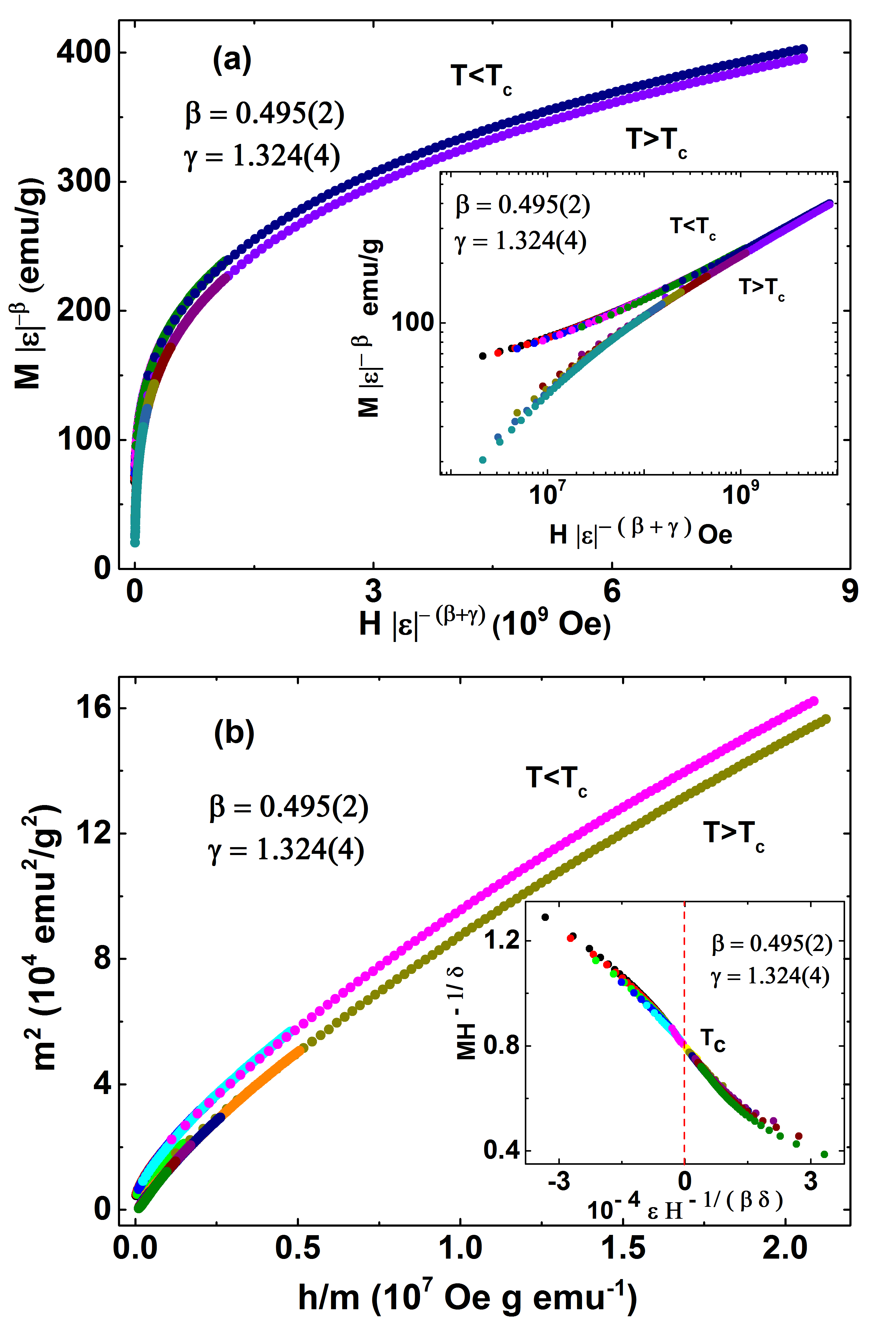}
\caption{(Color online) (a) Scaled magnetization as a function of the renormalized field for Co$_{2}$TiGe, indicating two separate branches of the scaling for isotherms below and above $T_{C}$. Inset shows the same plot in log-log scale. (b) The replotting of renormalized magnetization ($m$) and renormalized field ($h$) in the form of $m^{2}$ vs. $h/m$. Inset shows the rescaling of the $M$($H$) isotherms by $MH ^{-1/\delta}$ vs. $εH ^{-1/(\beta\delta)}$, where all the data collapse on a single curve following Eq. (8).}\label{rh}
\end{figure}

The values of the critical exponents $\beta$, $\gamma$, and $\delta$, obtained from various methods, are presented in the Table I along with the theoretically predicted values for different models. From the Table I, at a glance we see that the values of the exponents for the present system do not belong to any of the conventional universality class. Therefore, we must check whether these exponents can generate the scaling equation of state as given by Eq. (5). In Fig. 8(a), scaled-$m$ vs. scaled-$h$ has been plotted using the values of critical exponents $\beta$, $\gamma$, and $T_{C}$, calculated from the Kouvel-Fisher plot. The inset shows the same plot on log-log scale to expand the low-field region. From Fig. 8(a), it is clear that all the isotherms collapse onto two separate branches: one above $T_{C}$ and another below $T_{C}$. The reliability of the values of the exponents and  $T_{C}$ are further verified by plotting $m^2$ vs. $h/m$, as shown in Fig. 8(b). In this case, all the isotherms also fall on two independent branches of the scaling function. Furthermore, in the vicinity of the transition, the scaling equation of state can be rewritten in another form,
\begin{equation}\label{scaling}
\frac{H}{M^{\delta}}= k\left(\frac{\varepsilon}{H^{1/\beta}}\right),
\end{equation}
where $k$ is a regular scaling function. The scaling Eq. (8) implies that all the isotherms now collapse into a single curve. Following Eq. (8), we have plotted $MH^{-1/\delta}$ vs. $\varepsilon H^{-1/\beta \delta}$ in the inset of Fig. 8(b). The inset shows that all the experimental data fall on a single curve passing through $T_C$, which is located at zero on the horizontal axis of the plot. Therefore, the scaling behavior shown in Figs. 8(a) and 8(b), implies that all the interactions in the vicinity of $T_{C}$ get properly renormalized and the set of the critical exponents are same below and above the $T_{C}$ for the present compound.

The static critical exponents for Co$_{2}$TiGe, obtained from aforementioned rigorous methods, obey the scaling relation and the scaling equation of state. Therefore, these exponents seem to truly characterize the critical behavior in the compound. However, a comparison of these exponents with that predicted by different theoretical models for 3D system, confirms that estimated critical exponents for Co$_{2}$TiGe do not belong to any conventional universality class and fall between the values predicted by the 3D Heisenberg model and the mean-field theory. It should be mentioned that the asymptotic critical exponents, which are temperature independent in the asymptotic critical regime, truly characterize the critical behavior in a material belonging to a universality class. Usually, the asymptotic critical regime is $\mid$$\varepsilon$$\mid$ $\leq$10$^{-2}$ for homogeneous magnets and slightly larger for disordered and amorphous systems \cite{kaul}. The reduced temperature range, used to derive the critical exponents for Co$_{2}$TiGe, is 1.3$\times$10$^{-3}\leq\mid\varepsilon\mid\leq$1.4$\times$ 10$^{-2}$. Therefore, we have performed the above analysis in the asymptotic critical regime. Indeed, we noticed that the estimated effective critical exponents ($\beta_{eff}$ and $\gamma_{eff}$) almost remain same in the studied reduced temperature range\cite{pdbabu,kaul}. The observed discrepancy from the 3D Heisenberg values may result from two possibilities: (i) exchange interaction of extended-type, i.e., interactions beyond the nearest neighbor and (ii) magnetic inhomogeneity in the system.  Hence, it is necessary to understand the nature as well as the range of interaction in this material. Renormalization-group analysis has shown that the isotropic long-range dipolar interaction can drive a crossover towards mean-field exponents at a certain value $\varepsilon_{co}$ of the reduced temperature, which depends on the strength of the dipolar interactions \cite{kaul,aaharony,fisherme}. The analysis also shows that the new crossover exponents are only slightly different from the Heisenberg ones and shifted towards the mean-field values. The general theory of crossover phenomenon gives the crossover temperature as $\varepsilon_{co}$$\approx$ $g_{d}$$^{1/\phi_d}$, where $g_{d}$ is the strenght of the dipolar intearction and $\phi_{d}$ is the crossover exponent \cite{KRied}. $g_{d}$ can be estimated as $g_{d}$$\approx$$\frac{0.87}{T_C}$$\frac{p^2\theta}{v}$ \cite{aaharony,fisherme,KRied}, where $p=g \sqrt{S(S+1)}$, $\theta$ is a correction factor, and $v$ is the elementary cell volume in {\AA}$^3$. For the present material, $v$=196.93 {\AA}$^3$ and for spin $S$=1/2, $\theta$ can be taken as 0.68 \cite{KRied}. Using these parameters, we get $g_{d}$=2.426$\times$10$^{-5}$ for Co$_2$TiGe. Taking the crossover exponent $\phi_{d}$ as the Hesisenberg exponent $\gamma$=1.386, we estimate the crossover temperature to be $\varepsilon_{co}$=4.78$\times$10$^{-4}$ \cite{kaul,KRied}. Only for $\varepsilon$$<<$$\varepsilon_{co}$, the dipolar inteaction results in a new regime with exponents shifted towards mean-field values. Therefore, the isotropic dipolar picture is not consistent with the present scenario since the  estimated $\varepsilon_{co}$ for Co$_2$TiGe is well below the investigated asymptotic critical regime of 1.3$\times$10$^{-3}\leq\mid\varepsilon\mid\leq$1.4$\times$ 10$^{-2}$. However, the extended type of interaction can also arise from isotropic exchange interaction between spins involving itinerant electrons that decays spatially as $J(r)\propto1/r^{d+\sigma}$, where $d$ is the effective dimensionality of the spin interaction and $\sigma$ is a measure of the range of the exchange interaction \cite{Fisher}. In a 3D system with isotropic spins, the Heisenberg model is realized for $\sigma>$2 such that $J(r)$ decreases with $r$ faster than $r^{-5}$. Mean-field exponents are realized, when $J(r)\propto r^{-m}$ with $m<$4.5. For 3/2$\leq\sigma\leq$2, the system belongs to a different class with critical exponents that take intermediate values depending on $\sigma$. Given the dimensionality of lattice ($d$) and spin ($n$), $\sigma$ can be estimated using the renormalization-group approach which gives \cite{Fisher},
\begin{eqnarray}\nonumber
\gamma=1+\frac{4}{d}\left(\frac{n+2}{n+8}\right)\Delta\sigma+\frac{8(n+2)(n-4)}{d^{2}(n+8)^{2}}\\
\times\left\lbrace 1+\frac{2G(\frac{d}{2})(7n+20)}{(n-4)(n+8)}\right\rbrace \Delta\sigma^{2},
\end{eqnarray}
where $\Delta\sigma=(\sigma-\frac{d}{2})$ and $G(\frac{d}{2})=3-\frac{1}{4}(\frac{d}{2})^{2}$. Using Eq. (9), for ${d:n}={3:3}$ and $\sigma=1.88$, we have obtained the experimentally estimated value of $\gamma=1.32$. However, for $\sigma=1.88$, scaling relations\cite{Eugene} yield $\beta=0.393$ which is deviated from the 3D Heisenberg value towards the mean-field one but not as much deviated as observed experimentally (Table I).
The value of $\sigma$ suggests the extended nature of the exchange interaction, which could be due to the presence of RKKY-type long-range interaction between spins involving itinerant electron as argued by others and discussed before \cite{Trudel}. The values of the asymptotic critical exponents for uniform ferromagnet Ni (Table I) are also shifted towards the mean-field values and the shift is attributed to a crossover to the fixed-point corresponding to the isotropic long-range exchange interactions \cite{kaul}. But, in Co$_2$TiGe the large deviation in the value of the critical exponent $\beta$ from the 3D Heisenberg can not be explained only on the basis of extended nature of the exchange interaction. It should be mentioned that the mean-field like values of $\beta$ has been reported for several magnetically inhomogeneous systems \cite{sjpoon,nkhan,rvenka}. The amorphous FM Gd$_{80}$Au$_{20}$ also shows unusual values of the critical exponents (Table I), all being largely shifted towards the mean-field values, and are attributed to the dilution model for inhomogeneous FMs \cite{sjpoon,hmuller}. Magnetic inhomogeneities may lead to critical phenomena those could not be described on the basis of existing universality classes. However, unlike inhomogeneous FMs, the critical exponent $\gamma$ shows quite small shift towards the mean-field one for Co$_{2}$TiGe. Thus, all the obtained values of the critical exponents in Co$_{2}$TiGe can not be accounted for the magnetic inhomogeneity alone. Therefore, we believe that the observed unconventional critical exponents in Co$_{2}$TiGe comply with the complex nature of exchange interactions, involving both the short-range and beyond nearest neighbor spin-spin interactions, and also suggest some magnetic inhomogeneity in the system.

\section{Conclusion}

We elucidate the nature of exchange interactions in the half-mettalic full-Heusler ferromagnet Co$_2$TiGe by comprehensively studying the critical behavior in the vicinity of the ferromagnetic to paramagnetic phase transition at $T_{C}$=371.5 K. Although, magnetic Heusler alloys are regarded as an ideal local-moment system and their spin-spin interactions are modeled by a Heisenberg Hamiltonian, we observe a significant deviation in the values of critical exponents from the 3D Heisenberg towards the mean-field ones in Co$_2$TiGe. In particular, the value of the critical exponent $\beta$ is very close to the mean-field one. This implies diminished critical spin-fluctuation in the vicinity of the transition temperature, which could be due to the presence of exchange interactions beyond the nearest neighbor spin as well as magnetic inhomogeneity in the compound. The results are consistent with the recently argued fact that both the short-range direct exchange coupling between nearest neighbor and the indirect coupling between further neighbor spins mediated by long-range RKKY interaction, stabilize the long-range ferromagnetic ordering. The present study provides experimental evidence of the complex nature of exchange interactions in the a Co$_2$-based full Heusler ferromagnet, which does not belong to any known universality class predicted by theory.

\section{ACKNOWLEDGMENTS}

We thanks Mr. Arun Kumar Paul for his help during sample preparation and measurements.

\end{document}